\documentclass[aps,pre,showpacs,twocolumn]{revtex4}
\usepackage{amssymb}
\usepackage{graphicx}
\usepackage{dcolumn,amsmath,amsthm,amscd,amsfonts,amssymb,epsfig,graphics,graphicx,eucal}

\newcommand{\sn}{\mbox{sn}}

\begin{document}

\title{ Superfluidity of Bose-Einstein condensates in  toroidal traps with nonlinear lattices}
\author{ A. V. Yulin$^{1}$, Yu. V. Bludov$^2$, V. V. Konotop$^{1}$, V. Kuzmiak$^{3}$, M. Salerno$^4$ }
\affiliation{ $^1$Centro de F\'isica Te\'orica e Computacional and Departamento de F\'isica, Faculdade de Ci\^encias,
Universidade de Lisboa,   Avenida Professor
Gama Pinto 2, Lisboa 1649-003, Portugal
\\
$^2$  Centro de F\'{\i}sica, Universidade do Minho,
Campus de Gualtar, Braga 4710-057, Portugal
\\
$^3$ Institute for Photonics and Electronic of Czech Academy of Science, Chaberska 57, 182 51, Praha 8, Czech Republic
\\
$^4$Dipartimento di Fisica ``E.R. Caianiello'',
Universit\`a di Salerno, via ponte don Melillo, 84084 Fisciano (SA), Italy}

\date{\today}

\begin{abstract}
Superfluid properties of Bose-Einstein condensates (BEC) in toroidal quasi-one-dimensional traps are investigated in the presence of periodic scattering length modulations along the ring. The existence of several types of stable periodic waves, ranging from almost uniform to very fragmented chains of weakly interacting and equally spaced solitons, is demonstrated. We show that these waves may support persistent atomic currents and sound waves with spectra of  Bogoliubov type. Fragmented condensates can be viewed  as arrays of Josephson junctions and the current  as a BEC manifestation of the dc-Josephson effect. The influence  of linear defects on BEC superfluidity  has  been also investigated. We found that for subcritical velocities, linear defects that are static with respect to the lattice (while the condensate moves in respect to both the optical lattice and the defect)  preserve the BEC superfluidity.
\end{abstract}
\maketitle
\input{epsf.tex}
\epsfverbosetrue

\section{Introduction}
The possibility of particles flows in  multiply connected superfluid and superconducting systems  which can last  for  extraordinary long, in principle infinitely long, times
is certainly one of the most striking  manifestation of quantum mechanics at the macroscopic level~\cite{Leggett}. This fact,  well known for conventional  superconductors and superfluids,  has been recently observed  also in atomic Bose-Einstein condensates (BEC) confined in toroidal traps~\cite{exp2006,exp2007}, with persisting circulating currents. In addition to the fundamental interest of this  result  as a direct manifestation of the  BEC superfluidity~\cite{Pitaevskii}, matter waves in multiply connected geometries  also represent ideal systems for exploring nonlinear properties in the presence of nonzero density backgrounds. In this context, solitary matter waves in one-dimensional (1D) and 2D toroidal traps have been investigated  and the existence of a bandgap structure and of gap solitons have been reported in Ref.~\cite{Saito2004}. Possibility of management of  solitons, including acceleration and localization of matter waves, was explored in Ref.~\cite{BludKon}. The density distributions in a ring as a function of the relation between the healing length and the trap length was investigated in Ref.~\cite{Jackson2010}. Solitary waves  have also been studied for quasi 1D circular troughs of radially periodic 2D potentials~\cite{BMS2006}.

All these studies refer  to  the case of uniform  or periodic linear potentials present along the circumference of the trap. Moreover, several  experimental studies of BEC in linear optical lattices (OL) have shown the occurrence of interesting phenomena related to phase relationships between BEC wavefunctions in different sites, these including the  existence of Josephson currents \cite{Kasevich}, the proliferation of vortices in the Berezinskii-Kosterlitz-Thouless regime of BECs in 2D OLs  \cite{Cornell}, vortex nucleation  in rotating lattices of BECs \cite{Foot}, and so on.

On the other hand, it is known that interesting phenomena also arise when a linear optical lattice is combined with a periodic spatial modulation of the nonlinearity, also known as a nonlinear OL. This is particularly true for 1D settings  for which it has been shown that periodic spatial modulations of the scattering length, achieved with the help of the Feshbach resonance technique~\cite{Feshbach},
can induce long lived Bloch oscillations~\cite{SKB08}, Rabi oscillations \cite{BKS_PRA09}, and dynamical localization~\cite{BKS_EPL09} of gap solitons in the presence of accelerated linear periodic potentials. Periodic modulations of the nonlinearity have also been shown to be effective to generate solitons in 1D OLs \cite{malomed2005} and in random potentials \cite{garnier2005} to stabilize multidimensional solitons \cite{Luz2010} and to generate vortex rings in a highly controllable manner \cite{berloff} (for fresh review of this very active field of investigation see Ref.~\cite{KartMalTorn}).

We remark that toroidal traps are routinely created in laboratories with the aid of magnetic fields~\cite{circular_trap} and  nonlinear OLs have also been recently experimentally realized for BEC in linear settings~\cite{nonlin_latt_experim}. Methods to create trapping potentials for BECs by means of  rapidly moving laser beams  for a variety of geometries, including toroids, ring lattices and square lattices, have  also been recently developed \cite{boshier}.

To our knowledge, nonlinear OLs with a multiply connected geometry have not yet been realized. Considering  the  experimental rapid progresses in the field, however, it is reasonable to expect them to be soon  available. A nonlinear toroidal lattice could indeed be created by optically induced Feshbach resonances \cite{opticalFesh} using  an all-optical trap with an horizontal sheet beam  and a ring shaped vertical higher order Gauss-Laguerre  beam~\cite{Laguerre} whose intensity is periodically modulated along the ring due to phase interference (see, e.g., Ref.~\cite{ChaoKon}).  This setting would be especially effective in a 1D limit (for details of respective reduction see, e.g., Ref.\cite{BludKon}) and would permit to study the combined effects of nonlinearity and periodicity on the BEC superfluidity, a problem scarcely investigated in the literature.

The aim of this paper is to study superfluid properties of BECs in toroidal  traps with strong radial confinement and periodic modulations of the scattering length (nonlinearity) in the azimuthal direction in the framework of the mean field Gross-Pitaevskii equation (GPE).
In particular, we demonstrate the existence of several types of stable periodic waves, ranging from almost uniform to very fragmented chains of weakly interacting and equally spaced solitons, which can support   persistent currents and permit sound waves with interesting physical properties. The backgrounds that may exist in such traps can be classified in  terms of bifurcation diagrams in the parameter space $(\mu, \rho)$ , where $\mu$ is the chemical potential and $\rho$ is the density of the condensate. We analyze the bifurcations patterns and the stability properties of periodic matter waves  both with zero (stationary states)  and nonzero particle currents.
 The spectral properties of the linear excitations (sound waves) propagating
 in periodic condensates are also investigated and  the Bogoliubov type features  of the spectrum demonstrated.
In the presence of  a pronounced fragmentation of the condensate (i.e., in the soliton chain limit) we find that persistent currents of particles are expressed in the form  $\rho_{min} \sin \theta$, where $\rho_{min}$ is the minimum density of the condensate  and  $\theta$ is the  phase  difference of the order parameter of neighboring solitons. Quite interestingly, this expression is similar to the supercurrent expression of a Josephson junction in the zero voltage state (dc- Josephson effect), this suggesting  the interpretation of  fragmented condensates  as arrays of Josephson junctions and the persistent current as a BEC analog of the dc-Josephson effect~\cite{barone}.
 Superfluidity properties are also investigated in the presence of localized linear defects perturbing  the  nonlinear OL and  acting on the condensate as obstacle. Direct numerical simulations show that, in conformity with  the Landau criterion and in agreement with our sound wave analysis, the BEC superfluidity is preserved  in presence of  linear defects which  are static with respect to the nonlinear OL and that move in respect to the condensate  with subcritical velocities.

The paper is organized as follows. In Sec.~\ref{sec:model} we   formulate the mathematical model describing the condensate and analyze the ground states properties.   Families of periodic solitons in parameter space and their stability properties both for zero (stationary) and non zero carrying current states are discussed in in Secs. II A and II B, respectively.  The spectra of sound waves in the condensate and its  Bogoliubov structure are numerically calculated and discussed in Sec. III. In Sec. IV the superfluid stability properties of BEC  are investigated by direct numerical integrations of the GPE in the presence of a linear defect. Finally, in the Sec.V,  the main results of the paper  are summarized.

\section{Model equation}
\label{sec:model}
We consider the following dimensionless GP equation
\begin{eqnarray}
\label{GP}
    i\psi_t=-\psi_{xx}+U(x)|\psi|^2\psi + V(x)\psi,
\end{eqnarray}
where $U(x)$ is a periodical nonlinear potential  that accounts for  interatomic interactions  and $V(x)$ is a localized linear potential  modeling an obstacle  perturbing the flow of the ground state. Without loss of generality,  the period of the nonlinear lattice is chosen to be $\pi$, i.e. $U(x+\pi)=U(x)$ and  the circumference  of the trap is fixed to $L= M \pi$, with $M$ being a positive integer. The annular geometry obviously implies cyclic boundary conditions
\begin{equation}
\label{cyclic}
\psi(x,t)=\psi(x+L,t),
\end{equation}
 and permits the existence of current carrying states, with the  current, $J$,  defined as
\begin{eqnarray}
\label{J}
J =\frac{1}{2i}\left(\psi^{*}  \psi_x -\psi  \psi_x^{*}\right)
\end{eqnarray}
 (here  the asterisk denotes complex conjugation).  The properties of these states
are investigated in the following two subsections.

\subsection{Ground states with zero current }
\label{sec:J=0}
We start from the the analysis of the stationary states having zero current: $J=0$,  assuming that no linear defect is added to the system: $V(x)\equiv 0$.
In this case the order parameter has a constant phase, and can be searched in the form $\psi=\Psi(x)e^{-i\mu t}$, where $\mu$ is the chemical potential and  $\Psi$ is   real and  solves the stationary GP equation
\begin{eqnarray}
	\label{GP_stat}
    \mu\Psi=-\Psi_{xx}+U(x)\Psi^3.
\end{eqnarray}
We are particularly interested in nonlinear potentials changing the sign of the interaction, i.e. in the situations in which the  repulsive and attractive interactions alternate along the length $L$  of the trap.
We remark that for {\it a priori fixed} particular solutions $\Psi(x)$ it is  possible to construct exact potentials $U(x)$  for Eq. (\ref{GP_stat}) using the "inverse engineering" technique explained in Ref.\cite{BK}.  An example of such a potential is given by
\begin{equation}
\label{pot_example_2}
	U(x)=\frac{\kappa q^2}{4}\frac{\sn(qx,\kappa )(3\kappa\,\sn(qx, \kappa)-2)}{1+\kappa \sn(qx,\kappa)}, \quad q=\frac{4K(\kappa)}{\pi}
\end{equation}
with  $\sn$  denoting the standard Jacobi elliptic function and $K(\kappa)$ the elliptic integral.
Notice that $U(x)$   is parametrized by the elliptic modulus  $\kappa \in[0,1]$ and depending on the value of $\kappa$, it represents either a sequence of nonlinear single wells (for $0<\kappa< \sqrt{5/ 3}-1$)  or a series of nonlinear double wells (for $  \sqrt{5/ 3}-1<\kappa <1$). In this paper we shall restrict our study to the case of single-well potentials, only.
We also remark  that in spite of the apparent complexity of the potential (\ref{pot_example_2}), it can be approximated very well by a Fourier series with only few  harmonics. Indeed, even for the limiting value $\kappa= \sqrt{5/ 3}-1$ we obtain for the coefficients of the Fourier series $U(x)=\langle U\rangle +\sum_{k\neq 0}c_ke^{2ikx}$, where $\langle U\rangle \approx 0.239$, $|c_{\pm 1}|\approx 0.36$, $|c_{\pm 2}|\approx 0.12 $, $|c_{\pm 3}|\approx 0.016$,  $|c_{\pm 4} |\approx 0.001$, and so on. This implies that the suggested periodic modulation of the nonlinearity can be simulated with accuracy of about $5\%$ by employing only two laser beams with the wavelengths $\pi$ and $\pi/2$ in dimensionless units.

\begin{figure}[h]
\epsfig{file=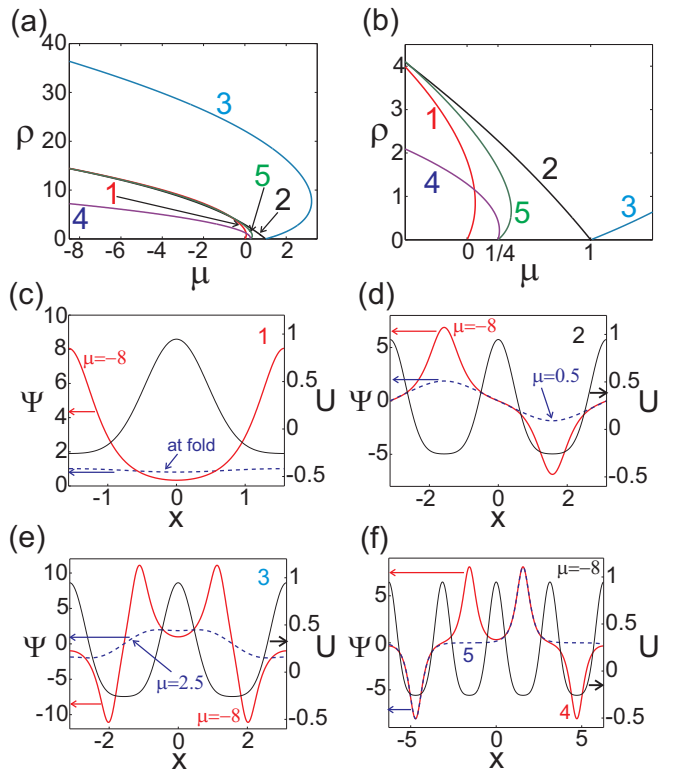,angle=0,width=\columnwidth}
\caption{(Color online) [(a) and (b)] Families of static solutions for the potential (\ref{pot_example_2}) centered at $x=\pi/2$ with $\kappa = 0.25$ [solid black lines in panels (c)-(f)]  bifurcating from  linear modes for the cases:} $L_0=\pi$ [red line (1)], $L_0=2\pi$ [black (2) and cyan (3) lines], and $L_0=4\pi$ [purple (4) and green (5) lines]. (c) The order parameter of the solutions is shown in panels (a) and (b) by red line (1). The solid red curve corresponds to the solution with $\mu=-8$ and the solution at the folding point $\mu=0.08$ is shown by the dashed blue line.   (d) The order parameter of the solutions is
shown in (a) and (b) by the black line (2) for $\mu=-8$ (solid red line) and for  $\mu=0.5$
(dashed blue line).
(e) The same as in panel (c)   but for the solutions corresponding to $\mu=2.5$ (dashed blue line) and $\mu=-8$  (solid red line) of the  family  shown by cyan line (3).
(f) The same as in panel (c)   but for the solution corresponding to $\mu=-8$
(dashed blue line)   of the  family  shown by the purple line (4) and  for the solution corresponding to $\mu=-8$  (solid red line)  of the  family  shown by the green line (5).
\label{fig2}
\vspace{0.1cm}
\end{figure}

An exact analytic solution of Eq. (\ref{GP_stat}) with $V=0$ and with the potential $U(x)$ taken  as in Eq.(\ref{pot_example_2}), can be  written as
\begin{eqnarray}
\label{sol_example_2}
	\Psi_a=\sqrt{1+\kappa \sn(qx, \kappa)},
\end{eqnarray}
with the corresponding chemical potential given by $\mu_a=\kappa^2q^2/4$. This solution is just a member of a family of periodic solutions which are parameterized  by the chemical potential $\mu$ and identified by the period $L_0$. Note that although  $L_0$ does not need to coincide with the period of the nonlinear potential $U(x)$, a relation $L=m L_0$ with $m$ integer, must be obviously satisfied. Since we have fixed $L=M\pi$ the set of possible   periods of the solutions is given by all $m$, which are integer divisors of $M$, i.e., $M/m$   is an integer and, respectively, $L_0=(M/m)\pi=p\pi$.
In Fig.~\ref{fig2} we illustrate the bifurcation diagrams  of these family solutions obtained numerically in the parameter space $(\mu,\rho)$, where $\displaystyle{\rho =\langle  |\Psi|^2\rangle }$ is the linear density of the condensate\footnote{Solution corresponding to quasi-periodical or non-periodical chains of solitons are also possible but they are  out of the scope of this paper} denoting the average of any  $L_0$-periodic function $f(x)$ as
$\langle f\rangle=\frac{1}{L_0}\int_{0}^{L_0}f(x)dx$, we have found that the potential (\ref{pot_example_2}) satisfies $\langle U\rangle>0$, i.e. the potential is repulsive in average.
In the present paper we restrict mainly to branches bifurcating from linear solutions for which the limiting transition $\rho \to 0$ is possible.
In this limit the ground state bifurcates from the uniform solution $\Psi_{c}= \sqrt{\mu}/\sqrt{\langle U\rangle} +{\cal O}(\mu^{3/2})$ and the respective branch has the period $\pi$.
The bifurcation points of the other solutions having linear limit can be designated by $\mu_{m}=4m^2/ M^2$. In each of such points, however there are two emergent modes: $\cos$- and $\sin$-like modes:

\begin{eqnarray}
\label{sin}
\Psi_{c}\approx
\sqrt{ \frac{\mu-\mu_{m}}{\frac{2 }{\pi }\int_{0}^{ \pi }U\left(\frac{Mx}{m}\right)\cos^4\left( 2 x\right)   dx}} \cos\left(\frac{2m}{M}x\right), \\
\label{cos}
\Psi_{s}\approx
\sqrt{ \frac{\mu-\mu_{m}}{\frac{2 }{\pi }\int_{0}^{ \pi }U\left(\frac{Mx}{m}\right)\sin^4\left( 2 x\right)   dx}} \sin\left(\frac{2m}{M}x\right).
\end{eqnarray}

Note that, formally, the ground state can be obtained from Eq.(\ref{sin}) by putting $m=0$.

From Figs.~\ref{fig2}(a) and \ref{fig2}(b) we observe that an increase of the  ground state density leads to the growth of the chemical potential, as is expected  for a BEC with repulsive interatomic interactions.
However, when the number of particles increases the nonlinear interactions become  important and the condensate becomes denser in the areas of attractive interactions.
At the point where the bifurcation diagram turns back, referred below as folding point, the chemical potential decreases [see the red line (1) in Figs.~\ref{fig2}(a) and \ref{fig2}(b)], and, for some average density, the chemical potential becomes negative. This happens because for larger densities most of the particles concentrate in the areas with attractive interactions and, therefore, the whole condensate behaves as having a negative scattering length. Mathematically, this is clear from the relation
\begin{eqnarray}
\label{H}
\mu \rho =  H = \langle\Psi_x^2\rangle+\langle U\Psi^4\rangle,
\end{eqnarray}
\begin{figure}[h]
\epsfig{file=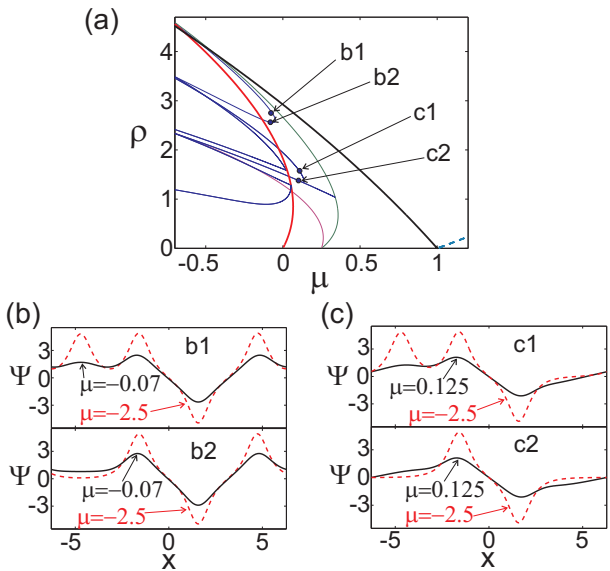,angle=0,width=8cm}
\caption{ (a) Solid lines show the dependencies of the linear density on the chemical potential  for the solutions with   $L_0 \leq 4\pi$  and having no more than one bright soliton in each area of attractive interaction.  The thick black and red lines correspond to the modes bifurcating from the linear solutions with the periods $\pi$ and $2\pi$ and, at large negative $\mu$, having one soliton in each area of the attractive interactions. The other two modes originating from linear ones are shown by the green and purple lines. Thin blue lines show the dependencies for the solutions that do not have a linear limit. The dashed thick cyan line shows a bifurcation curve which corresponds to the solutions originating from a linear one and having two solitons in each area of the attractive interactions.  (b) The field distributions for the solution corresponding to the points of the bifurcation curve denoted by b$1$ and b$2$ in panel(a) ($\mu=-0.07$) (black solid line). The red dashed lines show the solution with $\mu=-2.5$ and belonging to the same bifurcation branches.   (c) The same as in (b) but for the points marked in panel (a) by the circles c$1$ and c$2$; the chemical potentials in this case are $\mu=0.125$.
}
\label{fig3}
\vspace{0.1cm}
\end{figure}
from which we see that a negative $\mu$ implies $\langle U\Psi^4\rangle<0$ (note, that in the folding point $\mu_*$, where $d\mu/d\rho=0$ we have the relations $\mu_*=\frac H\rho=\frac{dH}{d\rho}$).
Thus a condensate  with sufficiently large density  can be considered as an array of in-phase bright matter solitons located at the areas with negative scattering length; see Fig.~\ref{fig2}(c).  Obviously, the interaction between the solitons decays exponentially as the density of the condensate increases, transforming in an array of in-phase matter solitons.

On the other hand, from the diagram bifurcating from the point $\mu=1$ [black (2) and   blue (3) lines in Figs.~\ref{fig2}(a) and \ref{fig2}(b)] we observe that the solution bifurcating from the sin mode [Eq.(\ref{sin})] at large negative $\mu$  can be seen as a chain of out-of-phase ($...0-\pi-0-\pi...$) bright solitons.
The solutions belonging to the second branch bifurcating from  Eq.(\ref{cos}) in the strongly nonlinear limit  also transform into a chain of pairs of bright solitons;  in this case there are two matter solitons in each domain of attractive interactions.

It is worth pointing out  that the solution families shown in Fig.~\ref{fig2} reveal a peculiar feature,
namely the kinetic energy of the condensate at the folding point is much larger than the potential energy $\langle\Psi_x^2\rangle \gg \langle U\Psi^4\rangle$.
We also note that not all periodical solutions bifurcate from linear modes: There are branches having no zero nonlinearity limit which is  however, expectable in view of the  analogous behavior of the localized modes~\cite{BludKon_sol}. Their bifurcation diagrams have folding points  or they merge with other solutions. In Fig.~\ref{fig3} the bifurcation diagrams are shown for all possible solutions which in the strongly nonlinear limit have no more than one soliton per period of the nonlinear potential and the period of these solutions is smaller than four periods of the nonlinear potential.

\subsection{Periodic matter waves with non zero current }
\label{sec:J.ne.0}

We now  consider periodic states carrying nonzero currents $J\neq 0$. Such states can be represented in the form
\begin{eqnarray}
\label{hydro_0}
\Psi (x)=\sqrt{\rho(x)}e^{i \theta(x)}  ,\qquad \theta(x)=\int v(x) dx,
\end{eqnarray}
where  $v(x)$ is the superfluid velocity and $\rho(x)$ the condensate density.
It is straightforward to show that such  solutions exist only if  $J=v(x)\rho(x)=const$ [see the definition, Eq.(\ref{J})], i.e. if $dJ/dx=0$, with the density $\rho$
satisfying the equation
\begin{eqnarray}
\label{hydro}
 \frac 1 2\frac{ d^2 \rho}{dx^2} -\frac {1}{4n}\left(\frac{d\rho}{dx}\right)^2  +\mu \rho -\frac{J^2}{ \rho}-  U(x)\rho^2=0
\end{eqnarray}
(note that zero-current states are characterized by densities which are strictly positive so the above equation is nonsingular).
In the following we consider $L_0$-periodic  density distributions
\begin{eqnarray}
\label{periodic_1}
\rho(x+L_0)=\rho(x)
\end{eqnarray}
of the form
$\rho(x)=\rho+\rho_1(x)$, with $\rho= \langle \rho(x)\rangle$ the mean density and $\rho_1(x)$  a  periodic variation of the density of period $L_0$. The condition $J=const$ then implies  that also the hydrodynamic velocity is of the form  $v(x)=v_0+v_1(x)$ with $v_0=\langle v\rangle$ the average superfluid velocity and $v_1$ a periodic function  of period  $L_0$.
The order parameter  $\Psi(x)$ has the form  of a nonlinear Bloch state
\begin{eqnarray}
\label{Bloch}
\Psi_{v_0}(x)=e^{i v_0 x}f_{v_0}(x)
\end{eqnarray}
with the mean velocity $v_0$ playing the role of quasimomentum and with $f_{v_0}(x)$ being a complex periodic function of period $L_0$. For the single valuedness or the wavefunction  $\Psi(x)$, the average superfluid velocity must be  quantized with respect to the length $L$ of the ring according to
$$
v_0=\frac{2\pi p}{L}= \frac{2  p}{M}
$$
where $p$ is an integer. We should remark that the analogy exploited above between superfluids in a ring and  Bloch electrons in crystals was first discussed by Bloch \cite{Bloch76} and  further developed in Ref.\cite{Saito2004}  for BEC in rotating traps.
Obviously, in the linear limit $f_{v_0}(x)$ is simply reduced to $\sqrt{\rho}$, and, taking into account, that in this limit $\rho\to 0$, one can neglect the spatial variations of the density and conclude that the bifurcation curves start  at $\mu=v_0^2$.
\begin{figure}[t]
\epsfig{file=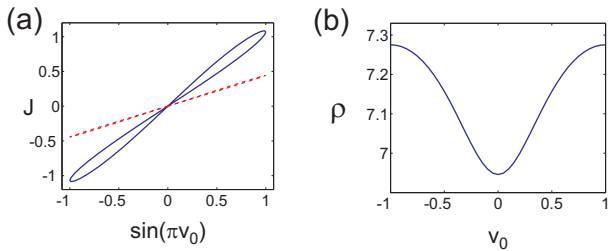,angle=0,width=8cm}
\caption{(a) The current {\it vs} the  average velocity for $\mu=-2$ (blue curve) and $\mu=-8$  (red curve).  (b) The average density  {\it vs} the  average velocity for $\mu=-2$.}
\label{fig4}
\vspace{0.1cm}
\end{figure}

It is of interest to investigate how the condensate with a nonzero current behaves when its average velocity $v_0$ is changed. This is done in Fig.~\ref{fig4} where the current and the average density versus  the average velocity are depicted for two fixed values of the  chemical potential.
Note that the current depends on the average superfluid velocity  periodically, acquiring maxima at semi-integer values $v_0=p/2$, while  at integer values of $v_0$ the current vanishes.
This is a direct consequence of the Bloch form (\ref{Bloch}) of the wavefunction   implying   that nondegenerate states with given $\mu$ and $v_0$ are equivalent to the state with the same chemical potential $\mu$ but with the velocity $v_0+p$, where $p$ is an integer.
Also note that at $v_0=0$ the solution in Fig.~\ref{fig4} is carrying zero current [as the symmetric state belonging to the bifurcation branch (1)] while  at $v_0=1$ the obtained periodic state is the asymmetric mode belonging to the branch (2) shown in Fig.~\ref{fig2} which is also a zero-current state.

Another interesting property to remark  is the quasi sinusoidal dependence of the current $J$ on $v_0$ for BEC wavefunctions fragmented as chains  of solitons on a nonzero background. The particle density of such condensates is depleted (enhanced) in correspondence with the regions of repulsive (attractive) interaction. Direct numerical simulations show that the stronger is the depletion of the condensate  in the repulsive  regions  the better the current is approximated by a sinusoidal law. In the limit of very strong depletion (i.e.,  for a chain of strongly localized and  weakly overlapping solitons), it becomes
\begin{equation}
J \approx \frac{1}{2}\sqrt{-\mu} \rho_{min} \sin \theta ,
\end{equation}
where  $\rho_{min} = -8\mu/ \max(U) \cdot \textmd{sech}^2(\sqrt{-\mu} \pi/2)$ denotes the minimum density of the condensate and  $\theta$ is the  phase  difference of the order parameter between neighboring solitons. We also note that in the case of strong localization the maximum condensate density corresponding to the soliton peaks, located at $x=\pi/2+\pi p$ where $p$ is an integer, is given by $\rho_{max} = 2\mu/ \min(U)$.
\begin{figure}[t]
\setlength{\epsfxsize}{3.3in} \centerline{\epsfbox{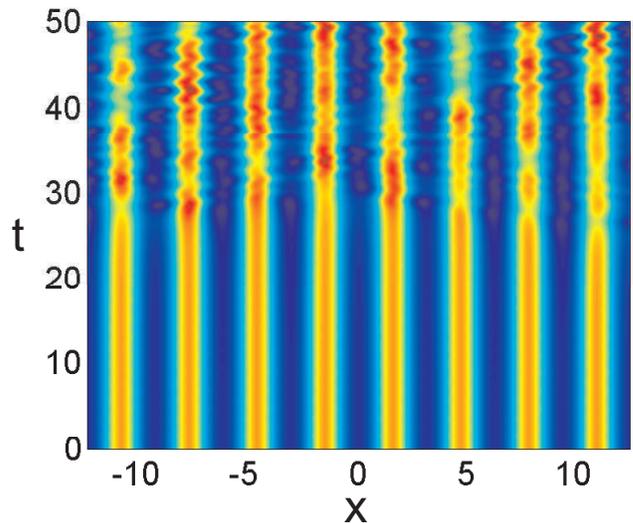}}\caption{ (Color
online) Temporal evolution of the BEC density for a current close to the critical value. Red color corresponds to the maximum density and blue color corresponds to the minimum density. The width of the window adjusted to show four periods of the initial field distribution. The parameters are $\mu=-2$, $v_0=\frac{8}{17}$, and $L=64\pi$.}%
\label{figure_n4}%
\end{figure}

Since the phase difference $\theta$ is constant for adjacent solitons, the above expression of $J$  is similar to  the  supercurrent flowing at  the zero voltage in a conventional superconductive   Josephson junction (e.g., to the dc-Josephson effect)~\cite{barone}. This  suggests  the interpretation of  the fragmented condensates  as an array of Josephson junctions and the persistent current as a BEC manifestation of the dc-Josephson effect~\cite{barone}.
The depleted areas between solitons (i.e., where the interaction is repulsive) play the role of insulating barriers of conventional  Josephson junctions. Following this analogy we can say that for stable symmetric ground states the depleted areas are the  analogous of  $0$-junctions, while for the stable asymmetric states the depleted areas correspond to $\pi$-junctions.
Also note that the role of the electric potential across  the junction is played by the  difference of chemical potentials between  neighboring solitons. Since the  fragmented state is a stationary state, it has a fixed chemical potential and, therefore,
the flow of particles across the depleted regions (supercurrent) in the analogy also occurs in to the presence of zero voltage.
Following this analogy, one can also expect  that the nonzero-current stationary solutions become unstable and disappear when the current approaches a critical value (critical current). This is shown in  Fig.~\ref{figure_n4} where the temporal evolution of the BEC density is depicted for a current very close to its critical calue. We see that in this case the condensate becomes very sensitive to perturbations and  develops instabilities that eventually lead to its destruction (see the next section). We also observe that the critical current decays exponentially with the height and width of the effective nonlinear barrier, although it never vanishes completely.

\section{Sound waves of periodic soliton}
\label{sec:spectra}

The time evolution of sound waves against nonlinear periodic backgrounds of the form (\ref{Bloch})  is governed by the linear equation
\begin{eqnarray}
\label{linear_2}
i \varphi_t =-   \varphi_{xx} -2i v_0     \varphi_{x} +
 (2U(x) |f_{v_0}|^2+v_0^2-\mu) \varphi
 \nonumber \\
 + U(x) f_{v_0}^2 \varphi^{*}.
\end{eqnarray}
Looking for a monochromatic excitation, $\varphi=A(x)e^{-i(\omega +i\gamma)  t}+B^*(x)e^{i(\omega -i\gamma)  t} $, where $\omega(k)$ and $\gamma(k)$ are real, and splitting exponents of the respective ODEs  one obtains the system of coupled equations with periodic coefficients:
\begin{eqnarray}
\label{linear_2c}
(\omega +i\gamma)A =-   A_{xx} -2i v_0 A_{x} +
 \nonumber \\
 (2U(x) |f_{v_0}|^2+v_0^2-\mu) A + U(x) f_{v_0}^2 B, \\
-(\omega +i\gamma)B =-   B_{xx} -2i v_0 B_{x} +
 \nonumber \\
 (2U(x) |f_{v_0}|^2+v_0^2-\mu) B + U(x) {f_{v_0}^*}^2 A.
\end{eqnarray}

These equations   allow  (due to the Floquet theorem)  for solutions of the form   $  \sim   \varphi_{k}(x) e^{i k x} $ with  $\varphi_{k}(x)$ a periodic solution of Eq. (\ref{linear_2})
[it is worth noting, that the period of the function $f_{v_0}$ is not necessarily equal to the period of the nonlinear potential  $U(x)$] because the coefficients in Eqs.(\ref{linear_2}) are $2U(x) |\psi_0|^2$ and $U(x) \psi_0^2$, with the background state $\psi_0$ a periodic function of $x$. [In this paper we restrict our consideration to periodic backgrounds only].

The eigenvalues  and the corresponding eigenfunctions can be found numerically: $\omega(k)$ yields the phase and the group velocities of the sound waves, while $\gamma(k)$ defines the stability of the ground state, the instability corresponding to positive $\gamma(k)$ (note that since the problem is Hamiltonian the real part $\gamma$ can appear only in pairs $\pm \gamma$).

\begin{figure}[ptb]
\setlength{\epsfxsize}{3.3in} \centerline{\epsfbox{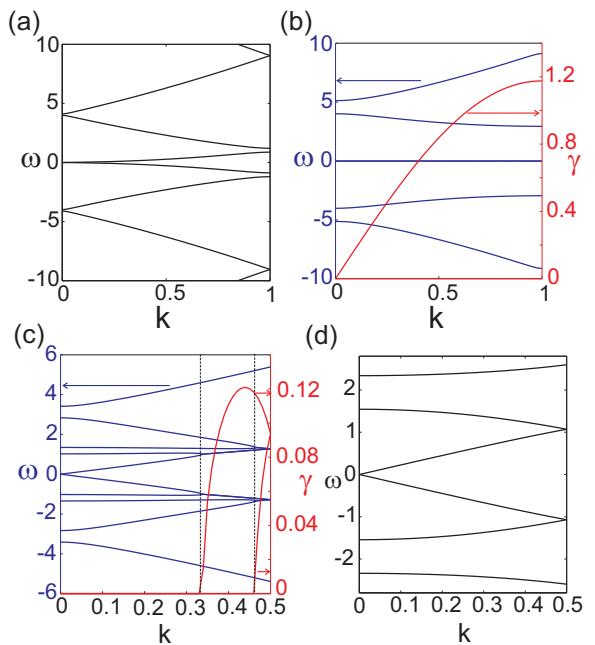}}\caption{ (Color
online) Sound wave spectra for the symmetric mode corresponding to curve 1  in Fig.~\ref{fig2}(b)  at $\mu=0.04$ (a) and at $\mu=-2$ (b) on the lower and upper branches of the bifurcation curve, respectively. Panels (c) and (d) refer to sound wave spectra for backgrounds modes taken on curve 2 of Fig.~\ref{fig2}(b)  at  chemical potentials  $\mu=0$ and $\mu =-2$, correspondingly. The black and blue curves correspond to  $\omega(k)$  (left vertical axis) and the red curves correspond to  $\gamma (k)$  (the right vertical axis).}%
\label{figure_n1}%
\end{figure}

Now let us turn to the stability of the backgrounds with zero current whose spectrum  is symmetric with respect to the transformation $k \rightarrow -k$. First, we consider the symmetric modes corresponding to the curve  1  in Fig.~\ref{fig2}(b). In this case the period of the solution coincides with the period of the nonlinear potential and the width of the Brillouin zone of the linearized problem is equal to $2$. The solutions belonging to the lower branch of the bifurcation diagram are stable. The typical spectrum of the linear excitation is shown in Fig.~\ref{figure_n1}(a). The spectrum is pure real and is of Bogoliubov type so $v_g=\partial_k \omega \neq 0$ at $k=0$. At the folding point the background becomes  unstable, a fact that is confirmed by our spectral analysis. In particular,  we find that the instability is generated by the merging of the lowest branches of the dispersion curve [i.e., the ones corresponding to lower values of $|\omega(k)|$)], as illustrated in Fig.~\ref{figure_n1}(b) for the case $\mu=-2$ for which branches are completely merged.

Next we consider the antisymmetric backgrounds corresponding to curve 2 in Fig.~\ref{fig2}(b). In this case the coefficients in the linearized Eq.(\ref{linear_2}) have the period $\pi$ equal to the period of the potential only if the state is described by a pure real function $\Psi$, i.e., if the background  has zero current. In the general case the period is equal to $2\pi$ and the width of the Brillouin zone is equal to $1$. The spectra of the linear excitations on antisymmetric  backgrounds are shown in Figs.~\ref{figure_n1}(c) and \ref{figure_n1}(d).

The spectra shown in Fig.~\ref{figure_n1} reveal Bogoliubov structure; however, they behave differently as far as the stability concerns. Specifically,  the states with relatively low particle density are unstable; see Fig.\ref{figure_n1}(c). It is seen that there are two pairs of merging modes, and the instability appears exactly at the wavevectors where two modes merge.
On the other hand, at some threshold density $\rho \approx 3.4$ (the corresponding chemical potential $\mu \approx -0.2$) of the particles the antisymmetric state becomes stable; see the spectrum of the stable state shown in Fig.~\ref{figure_n1}(d). The peculiarity of this state stems from the fact that the ground state is strongly depleted in the areas with repulsive nonlinear interatomic interaction. Therefore,  for large negative chemical potentials some excitations can be considered as motion of weakly interacting drops of the condensate and this interaction becomes exponentially weak for large negative $\mu$. It means that the eigenvalues appear in the spectrum with very small absolute values and the chain of the droplets becomes very flat so even relatively weak external action results in the significant motion of the mutual phases of the droplets. Therefore, in this case even relatively weak external perturbation can result in the strongly nonlinear dynamics and in the decay of the condensate. As a result, even relatively weak perturbations can destroy stability due to nonlinear effects, while the state remains linearly stable.

\begin{figure}[ptb]
\setlength{\epsfxsize}{3.3in}
\centerline{\epsfbox{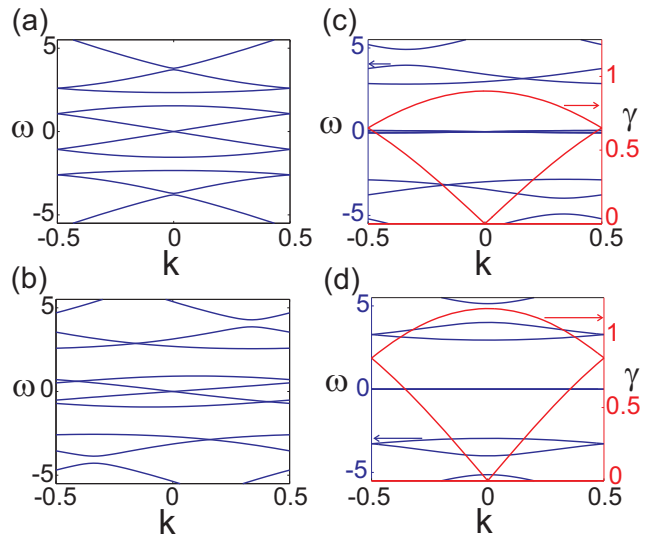}}
\caption{ (Color online)
Sound wave spectra for the state with $\mu=-2$ and the quasivelocities $v_0=0$ (a), $v_0=0.33$ (b),
$v_0=0.66$ (c) and $v_0=1$ (d). The antisymmetric state (a) and the symmetric one (d) modes
correspond curves 2 and 1 in Fig.~\ref{fig2}, respectively. Blue curves correspond to imaginary part $\omega$ of the
eigenvalue (left vertical axis) and the red curves to the real part $\gamma$ of the eigenvalues (right vertical axis).}
\label{figure_n2}%
\end{figure}

We now consider how the stability of the antisymmetric state changes when $v_0$ deviates from zero and the ground state has nonzero current. In general, the period of the coefficients is equal to $2\pi$; therefore, the width of the zone is equal to $1$. However, as mentioned above, at $v_0=0$ the period of the coefficients in the linearized equation is actually equal to $\pi$; -- see Fig.\ref{figure_n2}(a). For nonzero $v_0$ the period becomes equal to $2\pi$ and  gaps appear in the spectrum, as one can  see from Figs.~\ref{figure_n2}(b) and \ref{figure_n2}(c). One can also see  that the symmetry $k \rightarrow -k$ becomes broken for nonzero current. For relatively small values of $v_0$ the background remains stable and thus one can claim existence of stable ground states with nonzero current.

At some intermediate value of $v_0$ the modes start to merge, generating instability. The onset of the instability goes to the Josephson critical current with the depletion, but for finite depletion the instability appears at the current lower the critical one. The development of the instability is illustrated in Fig.~\ref{figure_n4}. We took the initial condition in the form of the numerically found stationary solution perturbed by weak noise and performed the modeling. One can see that the noise destroys the state during relatively short time. At $v_0=1$ the states transform into the symmetric mode considered above. As noted, this state is unstable. The gaps at the boundary of the zone become closed because the period of the coefficients becomes equal to $\pi$.

 Finally, we remark that the fact that the obtained spectra are of Bogoliubov type means that a perturbation moving (with subcritical velocity) with respect to the condensate does produce scattering  (a certain amount of scattering may be seen, however,  shortly after the introduction of the perturbation). This applies to the case of a perturbation that is resting with respect to the nonlinear potential, a  problem to be addressed  in the next section. The more general case of perturbations moving with respect to both the nonlinear lattice and the condensate will be considered elsewhere.

\section{Superfluidity of Bose-Einstein condensate}

It is well known that the scattering of the particles can be interpreted as Cherenkov synchronism between an obstacle and the propagating eigenmodes of the electron gas. When the system is described by the Schr\"{o}dinger equation, then a localized obstacle can be represented as localized linear potential in the equation. For the linear Schr\"{o}dinger equation,  with the dispersion of the elementary excitations parabolic,  the synchronism with a moving obstacle is inevitable unless the obstacle is resting with respect to the condensate. However, in the case of nonlinear Schr\"{o}dinger equation with an uniform repulsive (defocusing) nonlinearity the dispersion becomes linear at low $k$, implying  the existence of a critical velocity. The scattering occurs only if the velocity of the obstacle is larger than the critical  velocity.

\begin{figure}[ptb]
\setlength{\epsfxsize}{3.3in} \centerline{\epsfbox{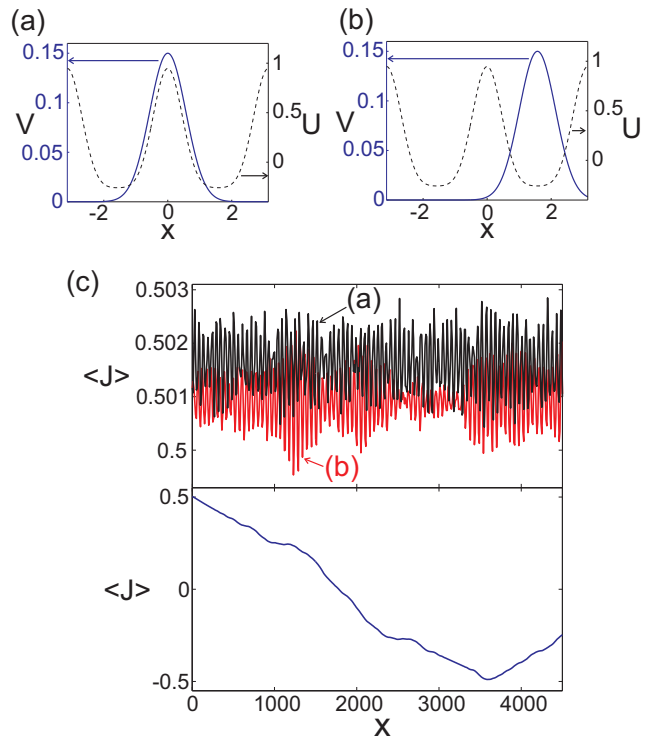}}
\caption{ (Color
online) The dynamics of BEC  in the presence of an obstacle described by the potential $V=a_0 \exp(-\frac{(x-x_0)^2}{w_0^2})$. Panels (a) and (b) show the relative positions of the nonlinear periodical potential and the linear potential modeling an obstacle, $x_0=0$ for panel (a) and $x_0=\pi/2$ for panel (b). The dynamics of the averaged current $\langle J\rangle $ is shown in the upper part of panel (c) for the relative position of the potentials shown in panels (a) (black line) and (b) (red line).  The dynamics of the averaged current for the case without nonlinear interatom interaction is shown in the bottom part of panel (c). The averaged density and the initial velocity of the particles are the same as for the case with nonlinear interactions. Parameters used are $\mu=-2$, $v_0=\frac{1}{8}$, $a=0.15$, $w_0=0.8$, and $L=64\pi$. Within the time scale shown in the figure the condensate makes about three rounds in the trap with  a duration of each round of  $512\pi$ in dimensionless time units.}
\label{figure_n3}
\end{figure}

In the previous section we have shown that there exist the spectra of linear excitations in Bose-Einstein condensates with spatially inhomogeneous nonlinear interaction of Bogoliubov kind with nonzero $v_g$ at $k=0$. Therefore, one can expect that at least an obstacle resting relatively to the nonlinear periodical potential will not lead the scattering of the Bose-Einstein condensate flowing in the system (the condensate is flowing relatively to both the periodical potential and the obstacle). To confirm this hypothesis we performed numerical simulations of the condensate flowing in the large annular system with periodic nonlinear potential. The results are summarized Fig.~\ref{figure_n3}.

One can see that for the stationary state the current is evenly distributed in space. If an obstacle leads to scattering, then the distribution of the current becomes inhomogeneous and, after some time, the averaged current tends to zero. We always calculated the distributions of currents assuming the initial condition the antisymmetric ground state discussed above and perturbing the equation by localized linear potential resting in respect to the periodical nonlinear potential. We carried out the simulations for very long time and no continuous  scattering has been observed; the dependence of the current averaged over space is shown in Fig.~\ref{figure_n3}(c).

The results of the numerical simulations indicate that introduction of the obstacle leads to very small perturbation of the current and the perturbation does not grow in time. This phenomenon is not sensitive to the relative position of the obstacle in respect to the nonlinear periodical potential; see Fig.~\ref{figure_n3}. We also verified  that the perturbation of the current  does not grow at arbitrary point of the system. For comparison, we show in  the bottom part of Fig.~\ref{figure_n3}(c) the dependence of the averaged current in the system with the same average particle density and the same initial current but without nonlinear interatom interaction. It is evident that the current oscillates and that the calculation time is considerably longer than the characteristic scattering time. The oscillation of the current in the linear system is, of course, periodic. Therefore, we can conclude that superfluidity of the BEC is observed in the system with periodic nonlinear potential.

\section{Conclusion}

We have explored the combined effect of nonlinearity and periodicity of nonlinear inter-atomic interaction in the toroidal traps where the scattering length is periodically modulated along the ring. First, we systematically studied periodic matter waves for both stationary states and states carrying nonzero current. The solutions found numerically have been parametrized by a chemical potential. We have shown for stationary states that when the nonlinear interaction changes its sign the condensate exists in the form of interacting drops localized in the areas with attractive interaction which rather counterintuitively implies the existence of the condensate with negative chemical potential despite of the dominating potential being repulsive. We also studied numerically the states with non-zero current and their relation to the stationary states. We have shown that the current depends periodically on the phase gradient and displays a strong fragmentation for large negative values of the chemical potential when the condensate is significantly depleted in the areas with repulsive interatomic interaction. We interpreted the fragmented condensates as Josephson $\pi$-junctions and the presence of the persistent current as a BEC analog of dc-Josephson effect which in contrast to a standard Josephson junction occurs in the system with and attractive effective inter-atomic potential.

Simultaneously, we studied sound waves propagating against the background. By inspecting the spectra of linear excitations on the ground states we investigated the stability of the condensate and we have shown that the spectrum are of Bogoliubov type. We found that the ground states consisting of interacting drops can be stable, which provides the strong evidence that these states and dc-Josephson effect can be experimentally observed.
On the basis of numerical simulations we have demonstrated that superfluidity of Bose-Einstein condensate may take place in the system with inhomogeneous nonlinear interatom interaction. In particular, we have shown that a localized defect does not lead to the scattering of the condensate when the defect is at rest with respect to the nonlinear potential while the condensate is moving in respect to both the periodic potential and the defect. We have observed this behavior in the systems with both an effective repulsive and attractive interatom interaction. The case when the defect is moving in respect to both the lattice and the condensate will be considered in a separate paper.

\bigskip

\section*{Acknowledgements}
AVY and VVK  were supported by the FCT (Portugal) under the grant PEst-OE/FIS/UI0618/2011.   AVY and VK were  supported by  the bilateral project.  VK was partially supported by the Ministery of Education, Youth and Sports of the Czech Republic  under the grant OC 09060. VVK was  partially supported by  the grant PIIF-GA-2009-236099 (NOMATOS) within the 7th European Community Framework Programme.
M. S. acknowledges support from Ministero dell' Istruzione, dell' Universit\'a e della Ricerca (MIUR) through the initiative "Programmi di Ricerca Scientifica di Rilevante Interesse Nazionale" (PRIN)-2008.


\begin{thebibliography}{99}

\bibitem{Leggett} A.J. Leggett, {\it Quantum Liquids}, Oxford Univ.Press, N.Y. (2006).

\bibitem{exp2006} M. F. Andersen, P. Clade, V. Natarajan, A. Vaziri, K. Helmerson,
and W. D. Phillips, Phys. Rev. Lett. {\bf 97}, 170406 (2006).

\bibitem{exp2007} C. Ryu, M. F. Andersen, P. Clade, V. Natarajan, K. Helmerson,
and W. D. Phillips, Phys. Rev. Lett. {\bf 99}, 260401 (2007).

\bibitem{Pitaevskii} L. P. Pitaevskii, "Bose-Einstein condensation: the odd nonlinearity quantum mechanics", in "Nonlinear Waves: Classical and Quantum Aspects", Eds. A. Kh. Abdullaev and V. V. Konotop (Kluwer Academic Publishers, 2004) pp. 175--192.

\bibitem{Saito2004} H. Saito and M. Ueda, Phys. Rev. Lett. {\bf 93}, 220402 (2004).

\bibitem{BludKon} Yu. V. Bludov and V. V. Konotop, Phys. Rev. A {\bf 75}, 053614 (2007).

\bibitem{Jackson2010} J. Smyrnakis,  M. Magiropoulos,  G.M. Kavoulakis, and A.D. Jackson, Phys. Rev. A {\bf 82}, 023604 (2010).

\bibitem{BMS2006} B.B. Baizakov, B. Malomed,  M. Salerno, Phys. Rev. E {\bf 74}, 066615 (2006).

\bibitem{Kasevich} B.P. Anderson and M.A. Kasevich, Science {\bf 282}, 1686 (1998).

\bibitem{Cornell} V. Schweikhard, S. Tung, and E. A. Cornell, Phys. Rev. Lett.  {\bf 99}, 030401 (2007).

\bibitem{Foot} R. A. Williams, S. Al-Assam, and C. J. Foot, Phys. Rev. Lett.  {\bf 104}, 050404 (2010).

\bibitem{Feshbach} P. O. Fedichev, Y. Kagan, G. V. Shlyapnikov,   and J. T. M. Walraven, Phys. Rev. Lett., {\bf 77}, 2913 (1996).

\bibitem{SKB08} M. Salerno, V.V. Konotop, and Yu.V. Bludov, Phys. Rev. Lett. {\bf 101}, 030405 (2008).

\bibitem{BKS_PRA09}Yu.V. Bludov, V. V. Konotop, and M. Salerno, Phys. Rev. A {\bf 80}, 023623 (2009).

\bibitem{BKS_EPL09}Yu.V. Bludov, V. V. Konotop, and M. Salerno, Europhys. Lett.  {\bf 87}, 20004 (2009); J. Phys. B: At. Mol. Opt. Phys. {\bf 42} 105302 (2009).

\bibitem{malomed2005}H. Sakaguchi and B. A. Malomed, Phys. Rev. E {\bf 72}, 046610 (2005);  Phys.Rev. E {\bf 73}, 026601 (2006).

\bibitem{garnier2005}F. Kh. Abdullaev and J. Garnier, Phys. Rev. A {\bf 72}, 061605 (2005).

\bibitem{Luz2010} H. L. F. da Luz, F. Kh. Abdullaev, A. Gammal, M.Salerno, and L. Tomio, Phys. Rev. A {\bf 82}, 043618 (2010).

\bibitem{berloff} N. G. Berloff and V. M. Perez-Garcia, arXiv:1006.4426v2.

\bibitem{KartMalTorn} Y. V. Kartashov, B. A. Malomed, and  L. Torner, Rev. Mod. Phys. {\bf 83}, 247 (2011)

\bibitem{circular_trap}S. Gupta et al., Phys. Rev. Lett. {\bf 95}, 143201 (2005); A. S.Arnold, C. S. Garvie, and E. Riis, Phys. Rev. A {\bf 73}, 041606(R) (2006).

\bibitem{nonlin_latt_experim} R. Yamazaki, S. Taie, S. Sugawa,  and Y. Takahashi, Phys. Rev. Lett. {\bf 105} 050405 (2010).

\bibitem{boshier}K. Henderson, C. Ryu, C. MacCormick1 and M.G. Boshier, New J. Phys. {\bf 11}, 043030 (2009).

\bibitem{opticalFesh}D.S. Petrov and G.V. Schlyapnikov, Phys. Rev. A {\bf 64}, 012706 (2001).

\bibitem{Laguerre}E. M. Wright, J. Arlt, and K. Dholakia, Phys. Rev. A {\bf 63},
013608 (2001); B. P. Anderson, K. Dholakia, and E. M. Wright, Phys. Rev. A {\bf 67}, 033601 (2003).

\bibitem{ChaoKon} Chao Hang and V. V. Konotop, Phys. Rev. A {\bf 83}, 053845 (2011)

\bibitem{barone}A. Barone, G. Paterno, {\it Physics and Applications
of the Josephson Effect}, John Wiley, N.Y. (1982).

\bibitem{BK} V. A. Brazhnyi and V. V. Konotop, Mod. Phys. Lett. B {\bf 18}, 627 (2004).


\bibitem{BludKon_sol} Yu. V. Bludov and V. V. Konotop, Phys. Rev. A {\bf 74}, 043616 (2006).

\bibitem{Bloch76} F. Bloch, Phys. Rev. A {\bf 7}, 2187 (1976).

\end{thebibliography}
\end{document}